# Structure of Extended Space

Tsipenyuk D.Yu.[*] and Andreev V.A.

General Physics Institute, Moscow, Russia

Abstract.

In work the generalization of Einstein's special theory of relativity on 5-dimentional space is considered, in which as fifth coordinates we consider the interval s of a particle. 5-dimentional vectors in this space are isotropic both for mass, and for massless of particles. In extended space, offered by the authors, there is a possibility in addition with usual Lorentz transformations, to enter two new transformations, therefore massive and massless of a particle can reversible converted to each other.

1. Introduction

We consider generalization of the Einstein's special theory of relativity on 5-dimentional space, to be exact speaking on (1 + 4) -dimensional space having the metric (+ - - - -) [1-5]. The physical basis for such generalization is that fact, that in spatial theory of relativity masses of particles are scalars and do not vary in case of the elastic interactions. However it is well known, that it is possible to consider a photon as a massless particle and to describe by a flat wave only in infinite blank space [7]. If the photon moving in a medium or it appears in the limited space, for example in a resonator or wave-guide, it acquires a nonzero mass.

We offered such generalization of special theory of relativity, which takes into account processes under which a mass of a particle m also would be a variable. As a mass of a particle m we, following the recommendations of the review [6], shall understand it a rest-mass, which is a Lorenz scalar. For this purpose first of all we shall construct the extension (1 + 3) - dimensional space of the Minkowski $M(T;\vec{X})$ on (1 + 4) - dimensional space $G(T;\vec{X},S)$. This space we will name as extended space.

Similarly it is possible to consider process of a modification of a mass and for other particles, for example, electrons assuming, that it depends on the environmental conditions and interactions.

Thus, it is represented natural to expand space of parameters describing a particle, with taking into accounts that it mass can vary during interactions.

Let's reduce simple analogy. The free particle is gone on direct, therefore to describe it a behavior, it is possible to limit (1 + 1) -dimensional space that consists from the time t and direction of movement x, as the remaining coordinates y and z remain constants. If the particle begins to interact with other objects, so that can leave with direct and begin to move also in a plane (YZ), it is not enough of such space already and it is necessary to expand up to (1 + 3) -dimensional space.

Precisely as well in our case, while the mass of a particle does not vary, it is possible to limit by the Minkowski space M (1,3), but if it begins to vary, the space M (1,3) should be expanded.

Attempts to unify of gravitation and the electromagnetisms have a large history.

The modern approaches to the given problem ascend to work of the F. Klein [8], in which he has shown, that the classical Hamilton mechanics can be presented as optics in space of the greater number of measurements.

Then Th. Kaluza has undertaken attempt to generalize the Einstein's theory of a gravitation to include in it electromagnetism [9]. He has offered to consider (1 + 4) -dimensional space with the metric dependent on potentials of an electromagnetic field. The Kaluza idea was advanced by O.Klein [10], H.Mandel [11] and V.Fock [12], the model, constructed by them, has received a title of the theory Kaluza-Klein. Was shown, that the trajectory of a charged particle has a kind of a geodesic line of zero length in 5-dimensional space.

---

[*] E-mail: tsip@kapella.gpi.ru

U. Rummer in the work on 5-optics [13] has offered to attribute to a new measurement dimensionality of an operation and to consider it periodic with period equal to the Plank constant $\hbar$. Let's mark, that in all these constructions the rest-mass of particles, in difference from developed in work [8-13] model of extended space, was considered as fixed magnitude.

The consequent development of the many-dimensional theories is described in the monography [14].

Close to an offered model of extended space is advanced in [16] approach. In this approach constructed (1 + 4) spaces, where as fifth the coordinates are offered to be used a mass (substance). However in this model (as and authors of [16] writes) it is impossible to construct, for example, energy-momentum tensor. In a model of extended space this defect is absent [1-3]. Offer in [16] to unify in one (1 + 4)-space time (t), space coordinates (x,y,z) and mass m (the substance) also is represented artificial. In a general theory of relativity for (1 + 3) Minkowski space (time – coordinate) conjugates (1 + 3) space (energy – momentum). In difference from [16] in our works [1-3] to the 5-th coordinate - interval in extended space (time – coordinate-interval) corresponds to 5-th a coordinate - mass in extended space (energy – momentum-mass).

The separate directions are many-dimensional constructions in theory of strings and superstrings [15].

The approach which we offered, essentially differs from all listed above models, though, conceptually intersects with some of them.

2. Structure of extended space

To each particle having mass m, in Minkowski space corresponds the hyperboloid that in a limiting case degenerating in a cone.

All points lying on a hyperboloid, have an identical interval — distance from to the beginning of the coordinates

$$s^2 = (ct)^2 - x^2 - y^2 - z^2. \qquad (1.1)$$

Case of modification mass of a particle corresponds to transition from one hyperboloid on other, i.e. the modification of an appropriate interval. For us is seems natural to select as an additional fifth coordinate interval s. Thus, we shall work in space with coordinates (t, x, y, z, s) and metric (+ - - - -). The considered objects lie in this space on a cone

$$(ct)^2 - x^2 - y^2 - z^2 - s^2 = 0. \qquad (1.2)$$

We shall designate it of G (1,4) or $G(T, \vec{X}, S)$.

The interval S is saved for usual Lorentz transformations in the Minkowski space $M(T; \vec{X})$ but varies under turns in extended space $G(T; \vec{X}, S)$. Thus, the Minkowski space $M(T; \vec{X})$ -is a cone in extended space $G(T; \vec{X}, S)$. In this space saved only

$$s^2 - (ct)^2 - x^2 - y^2 - z^2 = \text{const.} \qquad (1.3)$$

Usual (1 + 3)-dimensional cones and the hyperboloids arise as a cut of a surface (1.2) in space by hyperplanes $s = s_0$.

The gang of 5 numbers in each frame are set 5-vectors in space $G(T, \vec{X}, S)$

$$\bar{a} = (a_0, a_1, a_2, a_3, a_4) = (\tilde{a}, a_4) = (a_0, \vec{a}, a_4). \quad (1.4)$$

Here

$\bar{a}$ - is a 5-vector in the extended space $G(T, \vec{X}, S)$,
$\tilde{a}$ - is a 4-vector in Minkowski space $M(1,3)$,
$\vec{a}$ - is 3-vector in a Euclidean space $E(3)$.

In space of the Minkowski M (1,3) to each particles are compared an energy - momentum 4-vector [1].

$$\tilde{p} = (\frac{E}{c}, p_x, p_y, p_z). \quad (1.5)$$

In extended space $G(T, \vec{X}, S)$ we extend it up to a 5-vector

$$\bar{p} = (\frac{E}{c}, p_x, p_y, p_z, mc). \quad (1.6)$$

For free particles components of a vector (1.5) satisfy to the equation

$$E^2 - c^2 p_x^2 - c^2 p_y^2 - c^2 p_z^2 - m^2 c^4 = 0. \quad (1.7)$$

This is analog of a relation (1.2) in space $G'(E, \vec{P}, M)$, which conjugate to space $G(T, \vec{X}, S)$. The mass m is conjugate to the interval s. The transformations of the Lorentz change an energy of a particle E and it momentum $\vec{p}$, but leave by a constant a mass m. Transformations in extended space complementary to transformations of the Lorentz, change also mass of a particle, leaving a constant only form (1.8).

$$E^2 - c^2 p_X^2 - c^2 p_Y^2 - c^2 p_Z^2 - m^2 c^4 = const. \quad (1.8)$$

In difference from a usual relativistic mechanics, now we consider, that the mass of a particle m also is a variable. It should be understood in such a manner that the mass of a particle varies, when particle moves in area of space having a nonzero denseness of substance. As to in such areas slow down speed of light, we shall characterize its by their magnitude n - optical denseness. The parameter n links speed of light in a vacuum c with speed of light in a medium v: vn = c.

The set of magnitudes (1.6) defined 5-impulse, its components are saved, if the space $G(T; \vec{X}, S)$ is invariant in an appropriate direction. In particular, it fifth component $p_4$ have a sense of a mass, that does not vary, if the particle is gone so, that all time particle is in the field with constant density of substance or density of energy. This density of external substance (energy) can be interpreted, as a component of an external force affecting on a particle.

## 3. Vectors of free particles

We shall consider 5-vector of momentum of free particles in extended space $G(T,\vec{X},S)$ and study their transformations during the turns in this space. Necessary to take into account, that in a usual relativistic mechanics and field theory the mass of a particle is considered as constant and for particles with zero and nonzero rest-masses the different methods of the description are used. The massive particles are characterized by the mass m and velocity $\vec{v}$.

The particles with a zero mass (photons) are characterized by frequency $\omega$ and wavelength $\lambda$. This $\omega$ and $\lambda$ are connected with energy E and momentum $\vec{p}$ by such a way

$$E = \hbar\omega, \qquad \vec{p} = \frac{2\pi\hbar}{\lambda}\vec{k} \qquad (1.9)$$

To massive particle corresponds the 4-vector of energy - impulse $\tilde{p}$

$$\tilde{p} = (\frac{E}{c}, \vec{p}) = (\frac{mc}{\sqrt{1-\beta^2}}, \frac{m\vec{v}}{\sqrt{1-\beta^2}}), \qquad \beta^2 = \frac{v^2}{c^2}. \qquad (1.10)$$

To massless particle corresponds the 4-vector of energy – impulse $\tilde{p}$

$$\tilde{p} = (\frac{E}{c}, \vec{p}) = (\frac{\hbar\omega}{c}, \frac{2\pi\hbar}{\lambda}\vec{k}) = (\frac{\hbar\omega}{c}, \frac{\hbar\omega}{c}\vec{k}), \qquad \beta^2 = \frac{v^2}{c^2}. \qquad (1.11)$$

Here unit vector $\vec{k}$ define direction, in which photon move. Let's complete now of 4-vector (1.10), (1.11) up to 5-vectors. We attribute 5-vector for particle at rest with mass m

$$\bar{p} = (mc, \vec{0}, mc). \qquad (1.12)$$

It is possible to receive a particle moving with a velocity $\vec{v}$, if we pass in moving coordinate system. Now the vector (1.10) has a form

$$\bar{p} = (\frac{mc}{\sqrt{1-\beta^2}}, \frac{m\vec{v}}{\sqrt{1-\beta^2}}, mc). \qquad (1.13)$$

Similarly, the 4-vector (1.11) is generalized up to a 5-vector

$$\bar{p} = (\frac{\hbar\omega}{c}, \frac{2\pi\hbar}{\lambda}\vec{k}, 0). \qquad (1.14)$$

In case of transition in a moving coordinate system vector (1.14) does not change its form. The only change is the frequency magnitude.

$$\omega \rightarrow \omega' = \frac{\omega}{\sqrt{1-\beta^2}}. \qquad (1.15)$$

Thus, in blank space in a system coordinates at rest there are two different objects, with zero and nonzero masses, to which in space $G(T,\vec{X},S)$ corresponds of 5-vectors

$$(\frac{\hbar\omega}{c},\frac{\hbar\omega}{c},0). \qquad (1.16)$$

$$(mc,0,mc). \qquad (1.17)$$

## 4. Transformations in the extended space

Let's consider explicitly transformations in space $G(T,\vec{X},S)$.
All of them are reduced or to hyperbolic turns

$$x' = \frac{x + ct\tanh\varphi}{\sqrt{1-\tanh^2\varphi}} = x\cosh\varphi + ct\sinh\varphi. \qquad (1.18)$$

$$ct' = \frac{ct + x\tanh\varphi}{\sqrt{1-\tanh^2\varphi}} = ct\cosh\varphi + x\sinh\varphi.$$

Or to usual rotations in a plane

$$x' = x\cos\psi + y\sin\psi. \qquad (1.19)$$
$$y' = -x\sin\psi + y\cos\psi.$$

There are three types of transformations.

<u>1) Hyperbolic turns (1.18) in a plane (T, X).</u>

This is usual Lorentz turns which appropriated to transition in a moving coordinate system. Coordinate systems velocity v and turn angle $\varphi$ are connected by relations

$$\tanh\varphi = \beta = \frac{v}{c}, \qquad e^{2\varphi} = \frac{c+v}{c-v}. \qquad (1.20)$$

The magnitude $\varphi$ is an additive parameter, describing a turn in planes (T, X). If at first to execute a turn on an angle $\varphi_1$, and then on an angle $\varphi_2$, in an outcome the turn on an angle $\varphi_1 + \varphi_2$ will be received, i.e. in difference from velocities, the turn angles simply summarized. From the point of view of extended space it corresponds to an arbitrary movements in space - time of the Minkowski M (1,3) with a constant optical density n. In this turns of a vector of particles (1.16), (1.17) will be transformed as follows

$$(\frac{\hbar\omega}{c},\frac{\hbar\omega}{c},0) \to (\frac{\hbar\omega'}{c},\frac{\hbar\omega'}{c},0), \text{ here } \omega' = \omega e^{\varphi}, -\infty < \varphi < \infty. \qquad (1.21)$$

$$(mc,0,mc) \to (\frac{mc}{\sqrt{1-\tanh^2\varphi}},\frac{mc\tanh\varphi}{\sqrt{1-\tanh^2\varphi}},mc) = mc(\cosh\varphi,\sinh\varphi,1) \qquad (1.22)$$

For such transformation massless particles (photon) changes the energy and momentum, but does not acquire a nonzero mass. For massive particle the mass does not vary. Energy and momentum of the particles of both types vary, but 5-vector with them remain isotropic.

The transformations (1.18) save magnitude of an interval s, and length of vectors in the Minkowski space M (1,3) also they corresponds to areas of extended space with G (1,3, const). In this space they translate in themselves (1 + 3) -dimensional cones and hyperboloids. I.e. with their help it is possible to translate any point lying on a cone or a hyperboloid, in any other point lying on the same surface. But to pass from one such surface to other with the help of transformations (1.18) it is impossible.

2) We shall consider now turns in a plane (TS).

It is too hyperbolic transformations, which is looking like (1.18). Their physical sense is, that we don't make space movements, all of time we are in the same point, but the optical denseness in this point with change during the time.

Thus, in this case transformation (1.18) means transition to other moment of time and other optical density. All movements happen on 2-dimentinal cones and hyperboloids. As to no space movements are made, the momentums of particles should be saved.

In the case of transformations (1.18) photon vectors (1.16) will be transformed as follows

$$(\frac{\hbar\omega}{c},\frac{\hbar\omega}{c},0) \to (\frac{\hbar\omega}{c\sqrt{1-\tanh^2\theta}},\frac{\hbar\omega}{c},\frac{\hbar\omega\tanh\theta}{c\sqrt{1-\tanh^2\theta}}) = (\frac{\hbar\omega}{c}\cosh\theta,\frac{\hbar\omega}{c},\frac{\hbar\omega}{c}\sinh\theta) \quad (1.23)$$

As a result of such transformation originates particle with a mass

$$m = \frac{\hbar\omega\tanh\theta}{c^2\sqrt{1-\tanh^2\theta}} = \frac{\hbar\omega}{c^2}\sinh\theta \quad (1.24)$$

The transformation (1.23) is characterized by an additive parameter – angle . This angle we count from a light cone in to a positive direction of axes of time T. For such turns in upper half plane (t> 0) angles θ also vary from 0 up to ∞.
In the correspondence with the formula (1.24) in this case the photons have only positive mass. This outcome is represented completely natural, as the free photon vector (1.16) exists only in blank space with an optical density n = 1. The turn (1.23) increases an optical denseness of space and converts a photon into a particle with nonzero positive mass.

It is possible to calculate and velocity of such particle. For this purpose it is necessary to take advantage of the formula $v = c^2 p/E$. It gives

$$v = c\sqrt{1-\tanh^2\theta} = \frac{c}{\cosh\theta} \quad (1.25)$$

It is easy to check up, that, in spite of the fact that both mass and velocity of a particle vary, the momentum of particle remains constant. It is direct follows from relativistic expression for momentum of a particle with mass m, that moving with velocity v, and formulas (1.24), (1.25).

$$p = \frac{mv}{\sqrt{1-\frac{v^2}{c^2}}} = \frac{\hbar\omega}{c} \quad (1.26)$$

With the help of formulas (1.25) are possible to define connection between a turn angle and the magnitude of an optical density n.

$$n = \cosh\theta, \quad e^\theta = n \pm \sqrt{n^2 - 1} \qquad (1.27)$$

In the formula (1.27) it is necessary to keep both signs before a square root, as in both cases the right member is more than zero and has a physical sense. In that case revolved vector (1.23) acquires a kind

$$(\frac{\hbar\omega}{c}n, \frac{\hbar\omega}{c}, \frac{\hbar\omega}{c}\sqrt{n^2-1}). \qquad (1.28)$$

In the formula (1.28) we keep before a square root only sign "+", as to in the case of rotations in a half-plane (t> 0) masses of a photon is positive.
The massive vector (1.16) will be transformed as follows

$$(mc, 0, mc) \to (mce^\theta, 0, mce^\theta). \qquad (1.29)$$

In the case of such turn massive particle changes the mass and the energy, but momentum saved.

$$m \to me^\theta, \quad -\infty < \theta < \infty. \qquad (1.30)$$

That fact, that in the formula (1.27) both signs have a direct physical sense, means, that for such transformation from one particle with a mass m could originate two particles with different masses

$$m \to m_+ = me^{\theta_+}, \qquad (1.31)$$
$$m \to m_- = me^{\theta_-}. \qquad (1.32)$$

Thus, in the case of turns in a plane (TS) of a mass of particles with rest-masses, can be changed under two different laws (1.31) and (1.32).
In case of large n they have the following character of behavior

$$e^{\theta_+} = n + \sqrt{n^2-1} \to 2n - \frac{1}{2n}, \quad n \to \infty, \qquad (1.34)$$

$$e^{\theta_-} = n - \sqrt{n^2-1} \to \frac{1}{2n}, \quad n \to \infty, \qquad (1.35)$$

<u>3) The third type of a turn is a turn in a plane (XS).</u>

Actually space (X) is 3-dimensional with coordinates (x, y, z), but we select in it one direction (x) and we shall consider rotation in 2-dimensional space with coordinates (x, s).
All remaining transformations of a type 3) can be received, combining such rotations with usual 3-dimencional by space turns. The formulas (1.19) set a usual Euclidean turn. It corresponds to transition from space with one optical density in space with another optical density. In this case no temporary processes happens, all is considered in the same moment t. Therefore energy of particles is saved, and all processes, happening to them, are reduced to internal reorganizations. It could be understood so, that the particle, hitting in a more dense medium, is deformed and when abandoning it, restores the conditions. In this case does not happen exchange of energy or momentum between a medium and particle.

In the case of turn on an angle $\psi$ the photon vector (1.16) will be transformed under the law

$$(\frac{\hbar\omega}{c}, \frac{\hbar\omega}{c}, 0) \to (\frac{\hbar\omega}{c}, \frac{\hbar\omega}{c}\cos\psi, \frac{\hbar\omega}{c}\sin\psi) \ . \qquad (1.36)$$

Now the photon acquires a mass

$$m = \frac{\hbar\omega}{c^2}\sin\psi, \qquad (1.37)$$

and velocity

$$v = c\cos\psi, \qquad (1.38)$$

Using the formula (1.38), it is possible to connect a turn angle $\psi$ with magnitude of optical density n

$$n = \frac{1}{\cos\psi}, \qquad (1.39)$$

In this case the transformed photon vector (1.36) accepts a form of

$$(\frac{\hbar\omega}{c}, \frac{\hbar\omega}{cn}, \frac{\hbar\omega}{cn}\sqrt{n^2-1}) \ . \qquad (1.40)$$

The vector (1.16) massive particles will be transformed under the law

$$(mc, 0, mc) \to (mc, -mc\sin\psi, mc\cos\psi) = (mc, -\frac{mc}{n}\sqrt{n^2-1}, \frac{mc}{n}). \qquad (1.41)$$

Energy of the particle during such transformation is saved, but mass and impulse of the particle change

$$m \to m\cos\psi = \frac{m}{n}, \qquad (1.42)$$

$$0 \to -mc\sin\psi = -\frac{mc}{n}\sqrt{n^2-1}, \qquad (1.43)$$

The turns (1.19) can be made, on any angles, however formulas (1.36), (1.37) show, that, as we do not know what is a sense of negative masses of photons, it is necessary to limit by a range

$$0 \leq \psi \leq \pi. \qquad (1.44)$$

Now mass of a photon is always positive, and the possibility of emerging of a negative sign at photons momentum should be understood as a reflection of a photon from optically dense areas. The formulas (1.41) - (1.43), that describes transformation of a vector of a massive particle are similarly interpreted also. Such transformation physically corresponds to hit of a particle in area with a nonzero density of substance, which cause origination of the optical density n.

On a particle begins to act " the law of the Archimedes " and, in the correspondence with the formula (1.42), mass of the particle decreases and can even become negative, when the density of a particle will become less denseness of a medium. The formula (1.43) shows, that the "pushing out" force acts on such particle and transmit momentum, which always directed t opposite direction to movement.

Let's mark once again, that during all such transformations of an isotropic vector remain isotropic and, length of any vector from $G(T; \vec{X}, S)$ does not vary. If we shall consider space of the Minkowski M (1,3), in it with the help of transformations 1), 2), 3) it is possible to transfer any point in anyone.

## Conclusion

In the present work the generalization of Einstein's special theory of relativity on 5-dimentional space is considered, in which as fifth coordinates we consider the interval s of a particle. This generalization special theory of relativity on 5-dimensional space, allows constructing 5- vector $\bar{p} = (E/c, p_x, p_y, p_z, mc)$, where as fifth of coordinates the mass of a particle is selected.

All components 5-dimensional of a vector are connected by a well known formula : $E^2 = c^2 p_x^2 + c^2 p_y^2 + c^2 p_z^2 + m^2 c^4$. Within the framework of such approach there is no difference between massive and massless particles. In difference from 4-dimensional Minkowski space, in the entered extended space the 5-vectors become isotropic both for mass, and for massless particles.

As a result of magnification of dimensionality of space there are, except for the Lorentz transformation two another transformations of space (hyperbolic and Euclidean turns), which transfer massive particles into massless and visa versa.

During all of these transformations the isotropy of 5-vectors is not lost.